\documentstyle[12pt,aasms4]{article}

\begin{document}
\title{Measuring the galaxy power spectrum and scale-scale correlations
with multiresolution-decomposed covariance -- I. method}

\author{Li-Zhi Fang\footnote{fanglz@physics.arizona.edu}}

\affil{Department of Physics, University of Arizona, Tucson, AZ
85721}

\and

\author{Long-long Feng\footnote{fengll@physics.arizona.edu}}

\affil{Center for Astrophysics, University of Science and
Technology of China, Hefei, Anhui 230026, P.R. China\\ National
Astronomical Observatories, Chinese Academy of Science, \\
Chao-Yang District, Beijing, 100012}

\begin{abstract}

We present a method of measuring galaxy power spectrum based on
the multiresolution analysis of the discrete wavelet
transformation (DWT). Besides the technical advantages of the
computational feasibility for data sets with large volume and
complex geometry, the DWT scale-by-scale decomposition provides a
physical insight into the covariance matrix of the cosmic mass
field.  Since the DWT representation has strong capability of
suppressing the off-diagonal components of the covariance for
selfsimilar clustering, the DWT covariance for all popular models
of the cold dark matter cosmogony generally is diagonal, or
$j$(scale)-diagonal in the scale range, in which the second or
higher order scale-scale correlations are weak. In this range, the
DWT covariance gives a lossless estimation of the power spectrum,
which is equal to the corresponding Fourier power spectrum banded
with a logarithmical scaling. This DWT estimator is optimized in
the sense that the spatial resolution is adaptive automatically to
the perturbation wavelength to be studied. In the scale range, in
which the scale-scale correlation is significant, the accuracy of
a power spectrum detection depends on the scale-scale or band-band
correlations. In this case, for a precision measurements of the
power spectrum, or a precision confrontation of the observed power
spectrum with models, a measurement of the scale-scale or
band-band correlations is needed. We show that the DWT covariance
can be employed to measuring both the band-power spectrum and
second order scale-scale correlation.

We also present the DWT algorithm of the binning and Poisson
sampling with real observational data. We show that the so-called
alias effect appeared in usual binning schemes can exactly be
eliminated by the DWT binning. Since Poisson process possesses
diagonal covariance in the DWT representation, the Poisson
sampling and selection effects on the power spectrum and second order
scale-scale correlation detection are suppressed into minimum.
Moreover, the effect of the non-Gaussian features of the Poisson
sampling can also be calculated in this frame. The DWT method is
open, i.e. one can add further DWT algorithms on the
basic decomposition in order to estimate other effects on the
power spectrum detection, such as non-Gaussian correlations and
bias models.

\end{abstract}

\keywords{cosmology: theory - large-scale structure of the
universe }

\section{Introduction}

Measuring the galaxy power spectrum has been and is being a
central subject of the large scale structure study. Although the
power spectrum is only a second order statistical measure of the
deviations of a random field, $\delta({\bf x})$, of mass density
from homogeneity, it directly reflects the physical scales of the
processes that affect structure formation. Mathematically, the
positive definiteness of the power spectrum is useful for
constraining the parameter space in comparing predictions with
data. Since the ongoing and upcoming redshift surveys of galaxies
will provide data of galaxy distribution with highly improved
quality and a larger quantity, it also requests to develop the
methods of measuring the power spectrum more precise and
computationally efficient.

Different methods of the power spectrum measurements adopt
different representations, or decomposition of the covariance
$Cov=\langle \delta({\bf x})\delta({\bf x'}) \rangle$, where
$\langle ... \rangle$ stands for an ensemble average. For a
representation given by a set of basis functions $\psi_i({\bf x})$
(sometimes referred as weight function), the random field is
described by the variables
\begin{equation}
X_i=\int  \delta({\bf x}) \psi_i({\bf x})d{\bf x},
\end{equation}
and the covariance is given by $Cov_{ij}=\langle X_iX_j \rangle$.
If the covariance in this representation is exactly or
approximately diagonalized, the diagonal elements $\langle|X_i|^2
\rangle$ would be a fair estimate of the power spectrum, or
band-power spectrum. Thus, measuring power spectrum mathematically
is almost a synonym of diagonalizing the covariance of the density
field $\delta({\bf x})$, or calculating the eigenvalues of the
covariance matrix.

Traditionally, the Fourier decomposition, and then, the Fourier
power spectrum are the popular tool to analyze a cosmic density
field, because the Fourier transform retains the translation
invariance of a homogeneous and isotropic universe. However, the
observed sample given by redshift surveys are not translation
invariant due to the selection effect and irregular geometry of
the surveys. To effectively compare the predicted power spectrum
with the observed galaxy distributions, the basis functions of the
decomposition should be chosen to incorporate with the selection
effect, sampling, and complex geometry of the data. As a result,
various decompositions for measuring the galaxy power spectrum
have been proposed (Tegmark, et al. 1998 and reference therein).
An ideal estimator of the power spectrum should match the
following conditions
\begin{itemize}
\item  $X_i$s are independent from each other, i.e. the data is
       decomposed into mutually exclusive chunks;
\item  $X_i$s retains all the information of the original data, i.e. the
       decomposed chunks are collectively exhaustive;
\item  It is computationally feasible;
\item  It allows us to take account of the systematic effects, such as
       redshift distortion, evolution, morphology-dependence, galactic
       extinction etc.
\end{itemize}
These ideal estimators are believed to be information lossless,
i.e. retaining all information of the power spectrum in the
original data.

We will study, in this paper, the estimator based on the
multiscale decomposition, i.e. the discrete wavelet transform
(DWT) representation. The DWT power spectrum estimator has been
applied to measure the power spectrum from samples of the
Ly-$\alpha$ forests of QSO's absorption spectra (Pando \& Fang
1998a.) The result has demonstrated that the DWT power spectrum
estimator can match the conditions listed above, especially it is
very helpful to overcome the difficulties of complex geometry and
sampling. Within the framework of DWT, this paper will present a
general working scheme for extracting the statistical characters
from the observational data, in which the selection effect,
sampling and binning are addressed.

It has been recognized recently that the non-Gaussian behavior of
$X_i$ is substantial for a precise measurement of the power
spectrum. The accuracy of a power spectrum estimation is
significantly affected by the so-called power spectrum
correlations induced by non-linear clustering (Meiksin \& White,
1998, Scoccimarro, Zaldarriaga \& Hui 1999). The power spectrum
correlation is also found to be essential for recovering the
initial power spectrum by a Gaussianization of observed
distribution (Weinberg 1992, Narayanan \& Weinberg  1998, Feng \&
Fang 1999). Thus, beyond the conditions mentioned above for an
ideal power spectrum estimator, one should add one more
requirement that the power spectrum correlation caused by the
non-linear clustering and Poisson sampling are calculable. We will
show that the power spectrum correlations, or the scale-scale
correlations, can be calculated in the DWT analysis.

Moreover, for popular models of the cold dark matter cosmogony,
including the standard cold dark matter models (SCDM), open CDM
model (OCDM), and flat CDM (LCDM), the scale-scale correlations
have been found to be negligible on large scales, and the
non-local scale-scale correlations are also negligible even on
small scales (Fang, Deng \& Fang 2000). That is, the effect of the
power spectrum correlations is largely suppressed in the DWT
representation. We will show how to take the advantage of this
suppression for a scale-by-scale approach of measuring the power
spectrum.

The paper will be organized as follows. \S 2 gives a brief
description of the DWT decomposition of the covariance of density
random field. The physical meaning and mathematical properties of
the $j$ diagonal and $j$ off-diagonal components of the covariance
will also be discussed. In \S 3, an optimized band power spectrum
estimator based on the DWT $j$ diagonal covariance is proposed. In
addition, the scale-scale correlation extracting from the $j$
off-diagonal components of the covariance is investigated. This
correlation gives the scale range in which the power spectrum
obtained by the $j$ diagonalization are information lossless. We
then present the algorithm for estimating the DWT band power
spectrum from observed galaxy catalog. It includes the DWT binning
(\S 4), and the DWT technique of dealing with Poisson sampling and
selection (\S 5). The discussions and conclusions are given in \S
6. A brief introduction of the DWT analysis is given in Appendix.

\section{Covariance of density fluctuations in the DWT representation}

\subsection{DWT decomposition of density fields}

For the sake of simplicity, we analyze a 1-D density distribution
$\rho(x)$ in the range $0<x<L$, which is assumed to be a
stationary random field. The density contrast is defined by
$\delta(x)=(\rho(x)-\bar{\rho})/\bar{\rho}$, where $\bar{\rho}=
\langle \rho(x) \rangle$, and $\langle ...\rangle$ stands for
ensemble average. It would be straightforward to extend the most
results to 2-D and 3-D. Some specific problems related with higher
dimension extension will be discussed in \S 6. In addition, the
redshift distortion will not be taken into account in this paper.

To ensure a multiscale decomposition of $\delta(x)$ to be
information-lossless, the natural working scheme is to adopt
discrete wavelet transformation (DWT) within the framework of
multiresolution analysis (MRA). The mathematical construction of
MRA theory is briefly sketched in Appendix A.

Let $\delta^P(x)$ be the periodic extension of $\delta(x)$, i.e.,
$\delta^P(x)=\delta(x-[x/L]\cdot L)$, where $[\eta]$ denotes
integer part of $\eta$. From eq.(A36), the density contrast
$\delta^P(x)$ can be decomposed in term of orthonormal wavelet
basis
\begin{equation}
\delta^P(x) = \sum_{j=0}^{\infty} \sum_{l=-\infty}^{+\infty}
  \tilde{\epsilon}_{j,l} \psi_{j,l}(x),
\end{equation}
The wavelet function coefficient (WFC), $\tilde{\epsilon}_{j,l}$,
is given by the inner product of
\begin{equation}
\tilde{\epsilon}_{j,l}=\langle \psi_{j,l} | \delta \rangle \equiv
\int_{-\infty}^{\infty} \delta^P(x) \psi_{j,l}(x)dx.
\end{equation}
which describes the density fluctuation on scale $L/2^j$ at
position $lL/2^j$. The WFCs are the variables of the random field
in the DWT representation. The original distributions can be
exactly and unredundantly reconstructed from these decomposed
variables.

By using the periodized wavelet function defined by
\begin{equation}
\psi_{j,l}^P(x)=\left (\frac{2^j}{L}\right )^{1/2}
  \sum_{n=-\infty}^{\infty}\psi[2^j(\frac{x}{L}+n)-l].
\end{equation}
where $\psi$ is the basic wavelet function [eq.(A21)], eq.(1) becomes
\begin{equation}
\delta^P(x) = \sum_{j=0}^{\infty} \sum_{l=-0}^{2^j-1}
  \tilde{\epsilon}_{j,l} \psi^P_{j,l}(x),
\end{equation}
The WFC can then be computed by
\begin{equation}
\tilde{\epsilon}^P_{j,l}= \int_{0}^{L} \delta^P(x)
\psi_{j,l}^P(x)dx
\end{equation}
We will always use the periodized functions below, and drop the
superscript $P$.

Furthermore, $\psi_{j,l}(x)$ is admissible [eq.(A27)], which
implies that $\psi_{j,l}(x)$ has zero mean if it is integrable,
\begin{equation}
\int \psi_{j,l}(x) dx =0.
\end{equation}
It then follows from eq.(2) that
\begin{equation}
\langle \tilde{\epsilon}_{j,l} \rangle =0
\end{equation}

The Fourier decomposition of the field $\delta(x)$ is given by
\begin{equation}
\delta(x)=\sum_{n = - \infty}^{\infty} \delta_n e^{i2\pi nx/L},
\end{equation}
where $n$ is an integer, and the Fourier coefficients, $\delta_n$,
is
\begin{equation}
\delta_n = \langle  n|\delta \rangle \equiv \frac{1}{L}\int_0^{L}
\delta(x)e^{-i2\pi nx/L}dx,
\end{equation}

Since both the bases of the Fourier transform and the DWT are
orthogonal and complete in the space of 1-D functions with period
length $L$, we have
\begin{equation}
\sum_{j=0}^{\infty} \sum_{l=0}^{2^j-1} \langle n |\psi_{j,l}
\rangle \langle \psi_{j,l} |n' \rangle = \delta^K_{n,n'}
\end{equation}
where $\delta^K_{n,n'}$ is the Kronecker Delta function, and
$\langle n| \psi_{j,l} \rangle$ the Fourier transform of the
wavelet $\psi_{j,l}$ given by
\begin{equation}
\hat{\psi}_{j,l}(n)\equiv
 \langle n | \psi_{j,l} \rangle =
 \int_{0}^{L}\psi_{j,l}(x)e^{-i2\pi nx/L}dx.
\end{equation}
Considering the wavelet $\psi_{j,l}(x)$ is related to the basic
wavelet $\psi(\eta)$ by eq.(A11), eq.(12) can be rewritten as
\begin{equation}
\hat{\psi}_{j,l}(n)  =\left(\frac{ 2^{j}}{L}\right )^{-1/2}
\hat{\psi}(n/2^j) e^{-i2\pi nl/2^j},
\end{equation}
where $\hat{\psi}(n)$ is the Fourier transform of the basic
wavelet
\begin{equation}
\hat{\psi}(n)= \int_{0}^{L} \psi(\eta) e^{-i2\pi n\eta}d\eta.
\end{equation}

Substituting expansion (9) into eq.(6) yields
\begin{equation}
\tilde{\epsilon}_{j,l}=\sum_{n= -\infty}^{\infty} \delta_n
\int_{0}^{L} e^{i2\pi nx/L} \psi_{j,l}(x)dx = \sum_{n=
-\infty}^{\infty} \delta_n \hat{\psi}_{j,l}(-n).
\end{equation}
Similarly, inserting expansion (5) into eq.(10) we have
\begin{eqnarray}
\delta_n & = &  \frac{1}{L}\sum_{j=0}^{\infty} \sum_{l=0}^{2^j-1}
\tilde{\epsilon}_{j,l} \hat{\psi}_{j,l}(n) \\ \nonumber
 &  = & \sum_{j=0}^{\infty}  \sum_{l=0}^{2^j-1}
\left (\frac{1}{2^{j}L}\right )^{1/2} \tilde{\epsilon}_{j,l}
e^{-i2\pi nl/2^j} \hat{\psi}(n/2^j), \ \ \ \ \ \ \ \ \ n \neq 0.
\end{eqnarray}
Equations (15) and (16) show that both the Fourier variables
$\delta_n$ and the DWT variables $\tilde{\epsilon}_{j,l}$ are
complete.

However, the statistical properties of the Fourier mode $n$ and
the DWT mode $(j,l)$ are quite different. For a non-Gaussian field
consisting of randomly homogeneously distributed clumps with a
non-Gaussian probability distribution function(PDF), the one-point
distributions of the real and imaginary components of the Fourier
modes could be still Gaussian. That is because the Fourier modes
are subject to the central limit theorem of random fields (Adler
1981). Even though the non-Gaussian clumps are correlated, the
central limit theorem still holds if the two-point correlation
function of the clumps approaches zero fast sufficiently (Fan \&
Bardeen, 1995.)  Thus, the non-Gaussian information could be lost
in the Fourier representation if the phases of the Fourier
coefficients are missing.

On the other hand, the DWT basis doesn't suffer from the central
limit theorem. A key condition necessary for the central limit
theorem to hold is that the modulus of the decomposition basis are
less than $C/\sqrt L$, where $L$ is the size of the sample and $C$
is a constant (Ivanov \& Leonine 1989). The Fourier basis
obviously satisfy this condition because of $(1/\sqrt L) |\sin
2\pi nx/L| < C/\sqrt L$, where $C$ is independent of $x$ and $n$.
While the DWT basis is compactly supported (Appendix A), and its
modulus does not satisfy the condition $<C/\sqrt L$. Consequently,
for the non-Gaussian fields, the one-point distributions of the
Fourier variables $|\delta_n|$ could be Gaussian, while for the
DWT variable $\tilde{\epsilon}_{j,l}$, the one-point distributions
show non-Gaussian (Pando \& Fang 1998b.)

\subsection{The WFC covariance and DWT power spectrum}

In the DWT representation, the covariance $\langle
\delta(x)\delta(x')\rangle$ is expressed by a matrix
$\langle\tilde{\epsilon}_{j,l}\tilde{\epsilon}_{j',l'}\rangle$
with subscripts $(j,l);(j',l')$. The elements of $j=j'$, $l=l'$
will be called diagonals, while $j=j'$ called $j$ diagonals, and
$j \neq j'$ the $j$ off-diagonals.

The Parseval's theorem for the DWT decomposition is (Fang \& Thews
1998)
\begin{equation}
\frac{1}{L}\int_0^L |\delta(x)|^2 dx = \sum_{j= 0}^{\infty}
\frac{1}{L}\sum_{l=0}^{2^j-1} |\tilde{\epsilon}_{j,l}|^2,
\end{equation}
which implies that the power of perturbations can be divided into
modes, $(j,l)$. $|\tilde{\epsilon}_{j,l}|^2$ describes the power
of the mode $(j,l)$. One can then define the DWT power spectrum by
the diagonals of the covariance matrix, i.e.\footnote{The DWT
power spectrum, or called scalogram, has been extensively applied
in signal analysis (e.g. Mallat 1999.)}
\begin{equation}
P_{j,l} =\langle\tilde{\epsilon}_{j,l}^2\rangle.
\end{equation}

Since the random variables $\tilde{\epsilon}_{j,l}$ are complete,
one can define a Gaussian field $\delta(x)$ by requiring that all
the variables $\tilde{\epsilon}_{j,l}$ are distributed as a
Gaussian process with the covariance
\begin{equation}
\langle\tilde{\epsilon}_{j,l}\tilde{\epsilon}_{j',l'}\rangle =
P_{j,l}\delta_{j,j'}\delta_{l,l'},
\end{equation}
and the zero ensemble average of all higher order cumulants of
$\tilde{\epsilon}_{j,l}$. Thus, a Gaussian field is completely
described by its DWT power spectrum $P_{j,l}$. For a homogeneous
Gaussian field, the DWT power spectrum $P_{j,l}$ is
$l$-independent, i.e. $P_{j,l}=P_j$.

Using eqs.(15) and (16), the covariance in the Fourier and DWT
representations can be converted from one form to another by
\begin{equation}
\langle\hat{\delta}_n\hat{\delta}^{\dagger}_{n'}\rangle =
\sum_{j,j'=0}^{+\infty}\sum_{l=0}^{2^j-1}\sum_{l'=0}^{2^{j'}-1}
\langle\tilde{\epsilon}_{j,l}\tilde{\epsilon}_{j',l'}\rangle
\hat{\psi}_{j,l}(n)\hat{\psi}^{\dagger}_{j',l'}(n')
\end{equation}
and conversely
\begin{equation}
\langle\tilde{\epsilon}_{j,l}\tilde{\epsilon}_{j',l'}\rangle =
\sum_{n, n'=-\infty}^{+\infty}
\langle\hat{\delta}_n\hat{\delta}^{\dagger}_{n'}\rangle
\hat{\psi}_{j',l'}(n')\hat{\psi}^{\dagger}_{j,l}(n).
\end{equation}

Therefore, for a homogeneous Gaussian field given by the DWT power spectrum
$P_j$, eq. (20) implies
\begin{equation}
\langle\delta_n \delta_{n'}^{\dagger}\rangle = P(n)\delta_{n,n'},
\end{equation}
where
\begin{equation}
P(n) = \sum_{j=0}^{\infty}P_j
 \left |\hat{\psi} \left (\frac {n}{2^j}\right) \right |^2.
\end{equation}
In the derivation of eqs.(22), we used
\begin{equation}
\sum_{l=0}^{2^j-1}e^{-i2\pi(n-n')l/2^j}=\delta_{n,n'}.
\end{equation}
Eq.(22) shows that for a homogeneous Gaussian $P_j$, the Fourier
power spectrum $P(n)$ is uniquely determined by the DWT power
spectrum $P_j$.

However, the reversed relation doesn't exist, i.e. one {\it
cannot} show that the DWT covariance is given by eq.(19) with
$P_{j,l}=P_j$ if the Fourier covariance is given by eq.(22). This
indicates that the Fourier and WFC covariance are not equivalent.
For instance, fields consisting of homogeneously distributed
non-Gaussian clumps generally do not satisfy eq.(19) with a
$l$-independent $P_{j,l}$, but do so for eq.(22). That is, eq.(19)
with a $l$-independent $P_{j,l}$ places stronger constrains on the
random field than eq.(22), and therefore, eq.(22) will hold when
eq.(19) with a $l$-independent $P_{j,l}$ holds, but not generally
true for the converse.

\subsection{$j$ off-diagonals of the WFC covariance}

We now identify the physical meaning of the $j$ off-diagonal
components of the WFC covariance.

When the ``fair sample hypothesis" (Peebles 1980) holds, or
equivalently, the random field is ergodic, the $2^j$ WFCs
$\tilde{\epsilon}_{j,l}$, $l=0...2^j-1$, for a given $j$ can be
taken as $2^j$ independent measurements, because they are measured
by projecting onto the mutually orthogonal basis $\psi_{j,l}(x)$.
Accordingly, the $2^j$ WFCs form a statistical ensemble on the
scale $j$. This ensemble represents actually the one-point
distribution of the fluctuations of the DWT modes at a given scale
$j$. The average over $l$ is thus a fair estimation of the
ensemble average.

For a Gaussian field, these one-point distributions are Gaussian.
However, even if the one-point distributions for all $j$ are
Gaussian, the density field $\delta(x)$ could still be
non-Gaussian. That is simply due to the statistical properties of
the WFCs $\tilde{\epsilon}_{j,l}$ for indices $j$ and $l$ are
independent. It is easy to construct a density field $\delta(x)$
for which the WFCs $\tilde{\epsilon}_{j,l}$ are Poisson or
Gaussian in its one-point distribution with respect to $l$, while
highly non-Gaussian in terms of $j$  (Greiner, Lipa \& Carruthers
1995). A simple example is demonstrated as follows. Suppose the
one-point distribution of the 2$^j$ WFCs,
$\tilde{\epsilon}_{j,l}$, on a scale $j$, is Gaussian. If the WFCs
on the scale $j+1$ is incorporated with those on the scale $j$,
e.g.,
\begin{eqnarray}
\tilde{\epsilon}_{j+1,2l} & = & a\tilde{\epsilon}_{j,l}, \\
\nonumber \tilde{\epsilon}_{j+1,2l+1} & = &
b\tilde{\epsilon}_{j,l},
\end{eqnarray}
where $a$ and $b$ are arbitrary constants, the one-point
distribution of the 2$^{j+1}$ WFCs $\tilde{\epsilon}_{j+1,l}$ is
also Gaussian. However, the coherent structure given by eq.(25)
leads to a strong correlation between $\tilde{\epsilon}_{j+1,l}$
and $\tilde{\epsilon}_{j,l}$, i.e. the scale $j+1$ fluctuations
are always proportional to those on the scale $j$ at the same
position. This is a local scale-scale correlation. One can also
design non-local scale-scale correlation by
\begin{eqnarray}
\tilde{\epsilon}_{j+1,2l} & = & a\tilde{\epsilon}_{j,l+\Delta l},
\\ \nonumber \tilde{\epsilon}_{j+1,2l+1} & = &
b\tilde{\epsilon}_{j,l+\Delta l},
\end{eqnarray}
where $\Delta l = 1,2..$. Eq.(26) leads to a strong correlation
between the fluctuations on scales $j+1$ and $j$, but at two places
with distance $\Delta l$.

Hence, in terms of the DWT representation, a homogeneous Gaussian
field requires that (1) the one-point distributions of the WFCs
with respect to $l$ are Gaussian, and (2) the distributions of
WFCs with different $j$'s are uncorrelated, such as
\begin{equation}
\langle \tilde{\epsilon}_{j+1,l}\tilde{\epsilon}_{j,l'}\rangle =0.
\end{equation}

Correspondingly, in the Fourier representation, a Gaussian field
also has two requirements (1) the one-point distributions of the
amplitudes of the  Fourier mode $|\delta_n|$ are Gaussian; (2) the
phases of $\delta_n$ are random. Therefore, eq.(27) is the DWT
counterpart of the Fourier random phase. However, it is difficult,
or practically impossible, to capture the phase information of
each Fourier modes. The local scale-scale correlation is
overlooked with the Fourier covariance.

In summary, the $j$ off-diagonals of the WFC covariance provide
the information of the scale-scale correlation. This non-Gaussian
feature arises from mode-mode coupling of gravitational
clustering, and cannot be measured by the higher order cumulants
of the one-point distribution for a given scale $j$, rather, the
cross correlation between the different scales. The covariance of
a system without scale-scale correlation will be $j$-diagonal,
i.e.
\begin{equation}
\langle \tilde{\epsilon}_{j,l}\tilde{\epsilon}_{j',l'}\rangle =
\langle \tilde{\epsilon}_{j,l}\rangle \langle
\tilde{\epsilon}_{j',l'}\rangle =0,  \ \ \ j\neq j',
\end{equation}
where eq.(8) has been used at the last step.

\section{Statistical information extracting from the WFC covariance}

\subsection{$j$-diagonalization of the WFC covariance}

It has been known that the DWT is powerful for data compression. For very
wide types of stochastic clustering processes, the off-diagonal components
of the covariance are strongly suppressed in the DWT representation. This
suppression is especially efficient for selfsimilar clustering. For instance,
one can show analytically that the covariance in the DWT representation is
exactly diagonal for some popular hierarchical models of structure
formations, such as the block model and its variants (Meneveau \&
Sreenivasan 1987, Cole \& Kaiser 1988). In this respect, the DWT basis
represents the adequate normal coordinates. In other words, the DWT analysis
can be understood as a Proper Orthonormal Decomposition (POD), or a
Karhunen-Lo\`eve transformation (e.g. Aubry et al. 1988), in regard
to the second order correlations of these stochastic clustering processes.

For more realistic models and observed samples, the WFC covariance
is not fully diagonal, but mostly $j$-diagonal. In fact, this
character has been evident from the measurement of the fourth order
scale-scale correlation in the observational samples such as the Ly$\alpha$
forest lines (Pando et al. 1998), the transmitted flux of QSO
absorption spectrum (Feng \& Fang 1999) and the APM bright galaxy
catalog (Feng, Deng \& Fang 2000). A common conclusion is that the
scale-scale correlations are very weak, and negligible on large
scales, i.e. $\langle
\tilde{\epsilon}^2_{j,l}\tilde{\epsilon}^2_{j',l'}\rangle =
\langle \tilde{\epsilon}^2_{j,l}\rangle
\langle\tilde{\epsilon}^2_{j',l'}\rangle$ for $j\neq j'$ and $j,
j' \leq J_{ss}$, where $J_{ss}$ denotes the scale above which the
scale-scale correlation is not significant. It is also true for
the mass distributions and 2-D and 3-D mock catalog of galaxies in
the CDM family of models (Feng, Deng \& Fang 2000).  This result
indicates $\langle
\tilde{\epsilon}_{j,l}\tilde{\epsilon}_{j',l'}\rangle = \langle
\tilde{\epsilon}_{j,l}\rangle
\langle\tilde{\epsilon}_{j',l'}\rangle = 0$ for $j\neq j'$ and $j,
j' \leq J_{ss}$. Of course, the typical scale $J_{ss}$ relies on
the models or observational samples.

Therefore, on large spatial scales, $j \leq J_{ss}$, the WFC
covariance is already $j$-diagonal. Within this range, the covariance
matrix is decomposed into $j$ sub-matrices $\langle
\tilde{\epsilon}_{j,l}\tilde{\epsilon}_{j,l'}\rangle$. This guide
us to design the first statistics -- the DWT band-power spectrum.

\subsection{The DWT band-power spectrum}

Because the model-predicted power spectrum is currently expressed
in the Fourier representation, any statistical estimator designed
for measuring the power spectrum from real data should have simple
relation with the Fourier power spectrum.

Since we have only one realization of the cosmic mass field, no
ensemble is available for each mode $n$. One cannot measure the
Fourier power spectrum $P(n)$, as it is from the variance of the
amplitude $|\delta_n|$ of mode $n$. Generally, a power spectrum
estimator is to measure banded power spectrum as
\begin{equation}
P_j = \sum_n W_j(n)P(n),
\end{equation}
where $W_j(n)$ is a window function, which is localized in the $n$
(or Fourier) space. The problem that arises here is, what is the
criterion for a reasonable banding? and how to optimize the banded
power spectrum? The DWT representation provides a natural and
reasonable way for the banding.

As discussed in \S 2.2, for an ergodic field, the $2^j$ WFCs
$\tilde{\epsilon}_{j,l}$ at a given $j$ formed an one point
distribution of the fluctuations at the scale $j$. Therefore, the
DWT power spectrum at the scale $j$ can be defined as the variance
of the one-point distribution, i.e.,
\begin{equation}
P_j=\frac{1}{2^j}\sum_{l=0}^{2^j-1}
 (\tilde{\epsilon}_{j,l} - \langle \tilde{\epsilon}_{j,l} \rangle)^2.
\end{equation}
Because of the zero mean of WFC $\langle \tilde{\epsilon}_{j,l}
\rangle$, [eq.(8)]. $P_j$ can be written as, statistically,
\begin{equation}
P_j =\frac{1}{2^j}\sum_{l=0}^{2^j-1}
 |\tilde{\epsilon}_{j,l}|^2 = \frac{1}{2^j}\sum_{l=0}^{2^j-1}P_{j,l},
\end{equation}
which is an ergodicity-allowed spatial average of $P_{j,l}$, and
is usually referred as DWT power spectrum. As we will show below,
Eq.(31) gives an estimator of band-average Fourier power spectrum.

The DWT power spectrum eq.(31) is certainly less detailed than the
power spectrum $P(n)$ or $P_{j,l}$. However, the numbers $P_j$ are
probably the maximum of statistically valuable band-power spectrum
which can be extracted from one realization of an ergodic field.
The optimum of this banding can be seen via the phase space $\{x,\
k\}$, where the wavenumber $k= 2\pi n/L$. Generally a set of
orthogonal and complete basis of multiresolution analysis
decomposes the entire phase space into elements with different
shape, but their volume always satisfies the uncertainty relation,
$\Delta x \cdot \Delta k \ge 2 \pi$. The ordinary Fourier
transform is not a multiresolution decomposition, but always takes
highest resolution of $k$, i.e. $\Delta k \longrightarrow 0$, and
lowest resolution of $x$, $\Delta x \longrightarrow \infty$.

To apply the ergodicity, we chopped the survey volume $L$ into
pieces $\Delta x$. If $\Delta x$ is too large, or $L/\Delta x$ too
small, the ensemble contains few members, and thus there will be
larger vertical errors placed on the estimated power spectrum. In
order to minimize this error, we may make the size of chopped pieces
$\Delta x$ to be small. Correspondingly, the width of window function
$\Delta k= 2\pi/\Delta x$ will broaden, and the scale resolution will
be poor, i.e., there will be a large horizonal error bar placed on the
estimated power spectrum. Thus, the optimal chopping can be achieved by a
compromise between these two trade-off factors $L/\Delta x$ and
$\Delta k$. Generally, $1/\Delta x$ is proportional to the
resolvable wavenumber, i.e.
\begin{equation}
1/\Delta x \propto k.
\end{equation}
therefore, the optimized banding  $\Delta k \Delta x = 2\pi$
requires
\begin{equation}
\frac{\Delta k}{k} = \Delta \ln k \simeq 1.
\end{equation}
That is, the optimized banding is in logarithmic spacing. To
detect small scale fluctuations (larger wavenumber $k$), the size
of the pieces $\Delta x$ is chosen to be smaller. To detect large
scale fluctuations (smaller wavenumber), the size of the pieces
$\Delta x$ is chosen to be larger. The wavelets $\psi_{j,l}(x)$
is constructed by dilating (i.e. changing scale) of the
generating function by a factor $2^j$ (Appendix A). Therefore, we
have $\Delta \ln k \sim 1 $. In this sense, the DWT is an optimized
multiscale decomposition (Farge 1992). Because the set of wavelet
basis is complete, one cannot have more independent bands than
$P_j$.

Under the assumption of a homogeneous Gaussian field, the DWT
power spectrum eq.(31) can be rewritten as
\begin{equation}
P_j = \frac{1}{2^j} \sum_{n = - \infty}^{\infty}
 |\hat{\psi}(n/2^j)|^2 P(n).
\end{equation}
where eqs.(11), (21) and (22) have been used. Comparing with eq.(29),
clearly, $P_j$ is a band-averaged Fourier power spectrum with the
window function
\begin{equation}
W_j(n)=\frac{1}{2^j}|\hat{\psi}(n/2^j)|^2.
\end{equation}
Generally, the function $\hat{\psi}(n)$ is non-zero in two narrow
wavenumber ranges centered at $n=\pm  n_p$ with width $\Delta
n_p$. Therefore $P_j$ is the band spectrum centered at
\begin{equation}
\ln n_j = j \log 2 + \log n_p,
\end{equation}
with the band width as
\begin{equation}
\Delta \log n=\Delta n_p/n_p
\end{equation}
which stays constant logarithmically. Eqs.(36) and (37) show that
the countable data set $\{P_j, j=1,2 ...\}$ represents
scale-by-scale band-averaged Fourier power spectrum with the
logarithmic spacing of wavenumber. $P_j$ is completely determined
by the Fourier power spectrum, and therefore, it should be
effective for constraining the parameters contained in the Fourier
power spectrum.

The band-power spectrum (31) can also be written as, alternatively,
\begin{equation}
P_j  = \frac{1}{2^j} \hbox{tr}\  Cov^j_{l,l'}
\end{equation}
where the matrix $Cov^j_{l,l'}$ is the $j$ submatrix of the
covariance, i.e.
\begin{equation}
Cov^j_{l,l'}=\tilde{\epsilon}_{j,l}\tilde{\epsilon}_{j,l'}.
\end{equation}
Therefore, $P_j$'s exhaust all information of the $j$ diagonals of
the WFC covariance. Eq.(38) shows that we actually need not to
diagonalize each $j$ submatrix, as $P_j$ is given by the trace of
the $j$ submatrix.

\subsection{Scale-scale correlations in second and higher orders}

In the range of $j > J_{ss}$, the scale-scale correlations become
significant, the DWT covariance will no longer be diagonal or
$j$-diagonal.

In this scale range, we should do somewhat diagonalization of the
DWT covariance. However, the scale-scale correlation may lead to
large errors of the diagonalization, even the diagonalization
becomes impossible. Let us consider the example of the scale-scale
correlation given by eq.(25). In this case, the variable
$\tilde{\epsilon}_{j+1;l}$ actually is linearly dependent
on $\tilde{\epsilon}_{j,l+\Delta l}$, and therefore the matrix
$\langle \tilde{\epsilon}_{j+1,l}\tilde{\epsilon}_{j,l'}\rangle$
is singular. It cannot be diagonalized. For instance,
for scales $j= 1, 2$, the covariance matrix now is
\begin{equation}
\left(
\begin{array}{ccc}
\tilde{\epsilon}_{1,0} \tilde{\epsilon}_{1,0} &
\tilde{\epsilon}_{1,0} \tilde{\epsilon}_{2,0} &
\tilde{\epsilon}_{1,0} \tilde{\epsilon}_{2,1} \\
\tilde{\epsilon}_{2,0} \tilde{\epsilon}_{1,0} &
\tilde{\epsilon}_{2,0} \tilde{\epsilon}_{2,0} &
\tilde{\epsilon}_{2,0} \tilde{\epsilon}_{2,1} \\
\tilde{\epsilon}_{2,1} \tilde{\epsilon}_{1,0} &
\tilde{\epsilon}_{2,1} \tilde{\epsilon}_{2,0} &
\tilde{\epsilon}_{2,1} \tilde{\epsilon}_{2,1}
\end{array}
\right )
=
\tilde{\epsilon}^2_{1,0} \left(
\begin{array}{ccc}
1 & a & b \\ a & a^2 & ab \\ b & ab & b^2
\end{array}
\right )
\end{equation}
Obviously, this matrix cannot be diagonalized.

More seriously, if the matrix elements have some uncorrelated
errors due to measurements, i.e. $\tilde{\epsilon}_{j,l}
\tilde{\epsilon}_{j',l'} \pm \Delta\tilde{\epsilon}_{j,l,j'l'}$,
the matrix (40) looks diagonalizable. However in this case the
minors of the matrix are given by the errors
$\Delta\tilde{\epsilon}_{j,l,j'l'}$, and therefore, the
diagonalization will be largely contaminated by the errors.

This example indicates that when the scale-scale correlations
appear, the number of the independent variables, and then the
signal-to-noise ratio. will decrease. we should not extract
the statistical properties of the covariance by a diagonalization.

Fortunately, our ultimate goal is not the mathematical
diagonalization, but discrimination among physical models of the
structure formation. An alternative to the full diagonalization is
to take the following two measures: (1) Using the $j$-diagonals of each
$j$ to calculate the band-power spectrum $P_j$ [eq.(31)]; (2)
using the $j$ off-diagonals to calculate the second order scale-scale
correlations. The second order scale-scale correlations is defined
as
\begin{eqnarray}
C_{j,j'}(\Delta l) &  = & \frac{1}{2^j} \sum_{l=0}^{2^{j}-1}
\tilde{\epsilon}_{j,l}\tilde{\epsilon}_{j',l'}, \ \ \ j>j',\\
\nonumber
 l' & = & {\rm mod}[l/2^{j-j'}]+\Delta l.
\end{eqnarray}
Like the band-power spectrum [eqs.(30) and (31)], $C_{j,j'}(\Delta
l)$ is defined by an ergodicity-allowed average. $C_{j,j'}(\Delta
l)$ measures the second order correlation between fluctuations on
scale $j$ and $j'$ at positions $l$ and $l'$. Since cosmic density
field is homogeneous, the correlation depends only on the
difference between $l$ and $l'$, i.e. $\Delta l L/2^{j'}$. For an
initially Gaussian field, the scale-scale correlations are
developed during the non-linear evolution of the gravitational
clustering.

Now, we can use the two statistics $P_j$ and $C_{j,j'}$ to
discriminate among models. Actually, the two statistics
discrimination would be more worth than the full diagonalization.
For instance, the model-predicted galaxy power spectra on smaller
scales are generally degenerate with respect to cosmological
parameters, i.e. models with different cosmological parameters
can yield the same galaxy power spectrum. This is because one
always can choose the bias model parameters to fit the prediction
with the observations. Therefore, to remove the degeneracy, an
independent measure for constraining the bias models is necessary.
The scale-scale correlation is found to be sensitive to the bias
model (Feng, Deng \& Fang 2000). Thus, for model discrimination,
the $j$-diagonal power spectrum plus scale-scale correlation would
be more useful than a full-diagonalization.

In a word, in the scale range of $j>J_{ss}$, we will extract the
valid statistical information from the covariance by $P_j$ and
$C_{j,j'}(\Delta l)$.

It should be pointed out that even when all $C_{j,j'}(\Delta l)$
vanish, one cannot conclude that the system is scale-scale
uncorrelated. In other words, that a decomposition $X_i$ yields a
diagonal covariance doesn't mean that the modes $X_i$ are really
statistical uncorrelated. There are many clustering models which
have diagonal covariance, but mode-mode statistics are correlated
on higher orders (Greiner, Lipa \& Carruthers 1995.) A diagonal
decomposition means only that mode-mode is uncorrelated on second
order.

The higher order generalization of $C_{j,j'}(\Delta l)$ is
straightforward. For instance one can measure the fourth order
scale-scale correlations by
\begin{eqnarray}
C^2_{j,j'}(\Delta l) &  = & \frac{1}{2^j} \sum_{l=0}^{2^{j}-1}
\tilde{\epsilon}^2_{j,l}\tilde{\epsilon}^2_{j',l'}, \ \ \ j>j',\\
\nonumber
  l' & = & {\rm mod}[l/2^{j-j'}]+\Delta l.
\end{eqnarray}
This correlation $C^2_{j,j'}(\Delta l=0)$ is essentially the same
as the so called band-band correlation defined by
\begin{equation}
T = \frac{\langle P_{j} P_{j+1}\rangle}
   {\langle P_{j}\rangle\langle P_{j+1}\rangle}.
\end{equation}
It has been shown that the precision of the Fourier band-power
spectrum estimator depends on the band-band correlation $T$
(Meiksin \& White 1998.) In the DWT representation, we arrive at
the similar conclusion that when $C_{j,j'}(\Delta l)$ or
$C^2_{j,j'}(\Delta l)$ are non-zero, i.e. when the DWT covariance
is not $j$ diagonal, we should test models by both the band-power
spectrum and scale-scale correlations. For samples of large scale
structure, the scale-scale correlations
$C^2_{j,j'}(\Delta =0)$ has been found to be significant on scales less
about 10 $h^{-1}$ Mpc (Pando et al 1998, Feng, Deng \& Fang 2000.)

\section{The DWT algorithm of data binning}

In the following two sections, we will discuss the algorithm for
estimating the band power spectrum $P_j$ and scale-scale
correlations $C_{j,j'}(\Delta l)$ from galaxy redshift surveys,
and other samples of large scale structures.

If the position measurement is perfectly precise, the observed
galaxy distribution can be written as
\begin{equation}
\rho^g(x) = \sum_{i=1}^{N_g} w_i\delta^D(x-x_i),
\end{equation}
where $N_g$ is the total number of galaxies, $\{x_i\}$ the
position of the $i$-th galaxy, $0\leq x_i \leq L$, $w_i$ its
weight, and $\delta^D$ is the Dirac-$\delta$ function. However,
the position measurement has error due to finite spatial resolution, and
therefore, the distribution usually is somewhat given by a binned
histogram.

The binning is performed by a convolution of the data with a binning
function $W(x)$ as
\begin{equation}
\tilde{\rho^g}(x) = \Pi(x)\int W( x-x') \rho^g(x'),
dx'
\end{equation}
in which $\Pi(x)$ is the sampling function defined as
$\Pi(x)=\sum_{l}\delta^D(x-lL/2^j)$, where $l$ labels the $l$-th bin.
Obviously, the mesh-defined density distribution is given by
$\tilde{\rho^g}(x)= \sum_{l}\rho^g_l\delta^D(x-lL/2^j)$,
where $\rho^g_l=\int W(lL/2^j -x') \rho^g(x')dx'$ is a mass assignment
at the $l$-th bin.

It is well known that the binning eq.(45) will result in spurious
features of the Fourier power spectrum on scale around the Nyquist
frequency of the FFT grid (e.g. Jing 1992, Percival \& Walden
1993, Baugh \& Efstathiou 1994). Mathematically, eq.(45) implies a
decomposition by the weight function $W(x)$. In other word,
$W(lL/2^j -x')$ are playing the role of a scaling functions (or sampling
function.) If the scaling functions are orthogonal and complete, the
one cannot recovered the original field without distortion. This
may cause some spurious features, such as the aliasing effect in the
FFT. In the DWT analysis, the binning or sampling are always done
by an orthogonal and complete decomposition, one can expected that
the spurious features and false correlations can be completely avoided.

\subsection{Binning with wavelets}

The WFCs $\tilde{\epsilon}_{j,l}$ are assigned at regular grids
$l=0...2^{j-1}$. It is actually a binning of data. In this case,
the binning is automatically realized by the orthogonal projection
onto wavelet space, and no extra weight function is required. In
result, the contamination due to the sampling error is naturally
eliminated.

With eq.(6), one can directly calculate the WFCs of the galaxy
distribution (44) by
\begin{equation}
\tilde{\epsilon}^g_{j,l} = \sum_{i=1}^{N_g} w_i\psi_{j,l}(x_i).
\end{equation}
The errors of $\tilde{\epsilon}^g_{j,l}$ can also be calculated
from the errors of $x_i$.

Since we used the periodized distribution $\delta(x)$ in eq.(6),
the discontinuity between the data at two boundaries may introduce
false coefficients. Yet, this possible false signal is only
related to boundaries. One can expected that this false
coefficients will not be important for detecting power spectrum on
scales much less than $L$. This boundary effect has been tested
numerically by using simulated samples over a finite length
divided in 512 bins with two different boundary conditions (A)
periodic boundary conditions; (B) zero padding. The results show
that the spectrum can be correctly reconstructed by the DWT
regardless of the boundary conditions on scales equal to and less
than 64 bins (Pando \& Fang 1998).

Note has to be taken of the difference between usual mass
assignment and the DWT projection (46). In the former, the mass
assignment is given by partitioning the mass on the grids
according to the binning function $W(x)$, and the binning data are
the mesh-defined densities. Whereas for the DWT projection, the
binning data, i.e. the WFCs $\tilde{\epsilon}^g_{j,l}$ are not the
mesh-defined densities, but the fluctuations on scale $j$ at
position $l$, which is obviously not positive-definite.

\subsection{Binning with scaling functions}

In the DWT analysis, the mass assignment is realized by the
scaling function $\phi_{j,l}(x)$ [eq.(A30)]. Besides the
orthogonality eqs.(A33) and (A34), the basic scaling function
$\phi(\eta)$ (which is not yet periodized!) satisfies the
so-called ``partition of unity'' as (Daubechies 1992)
\begin{equation}
\sum_{l=-\infty}^{\infty} \phi(\eta-l)=1.
\end{equation}
One can also define the periodized scaling function as
\begin{equation}
\phi_{j,l}^P(x)=\left (\frac{2^j}{L}\right )^{1/2}
  \sum_{n=-\infty}^{\infty}\phi[2^j(\frac{x}{L}+n)-l].
\end{equation}
Thus, eq.(47) can be rewritten as
\begin{equation}
\sum_{l=0}^{2^j-1} \frac{L}{2^j}\phi^P_{j,l}(x)=1
\end{equation}
We will only use the periodized scaling function below, and drop
the superscript $P$.

With the periodized scaling function, the eqs.(A39) - (A41) give
\begin{equation}
\rho(x) =\rho^J(x)+
 \sum_{j=J}^{\infty}\sum_{l=0}^{2^j-1}\tilde{\epsilon}_{j,l}\psi_{j,l}(x),
\end{equation}
where
\begin{equation}
\rho^J(x)=\sum_{l=0}^{2^J-1}\epsilon_{J,l}\phi_{J,l}(x).
\end{equation}
The scaling function coefficients (SFCs) $\epsilon_{J,l}$ is given
by
\begin{equation}
\epsilon_{J,l} =
  \int_{0}^{L} \rho(x)\phi_{J,l}(x) dx
\end{equation}

Subjecting the distribution (44) to the transform eq.(50), we have
\begin{equation}
\rho^g(x) =\sum_{l=0}^{2^J-1}\epsilon^g_{J,l}\phi_{J,l}(x)+
\sum_{j=J}^{\infty}\sum_{l=0}^{2^j-1}\tilde{\epsilon}^g_{j,l}\psi_{j,l}(x),
\end{equation}
where
\begin{equation}
\epsilon^g_{J,l} = \sum_{i=1}^{N_g}w_i\phi_{J,l}(x_i).
\end{equation}
Using eqs.(44) and (54), eq.(49) yields
\begin{equation}
\sum_{l=0}^{2^j-1} \frac{L}{2^j}\epsilon^g_{j,l}
   = \sum_{i=1}^{N_g}w_i.
\end{equation}
This shows that the $i$-th galaxy is assigned onto grid $l$ by
number $(L/2^j)w_i\phi_{J,l}(x_i)$. Therefore, the SFC
$(L/2^j)\epsilon^g_{j,l}$ is the mass assignment of $\rho^g(x)$.

\subsection{The DWT binning and FFT}

Given a galaxy distribution eq.(44), its Fourier transform is
evaluated by the trigonometric summation
\begin{equation}
\hat{\rho}^g(n)=\sum_{i=1}^{N_g}w_ie^{i 2\pi nx_i/L},
\end{equation}
and the power spectrum is $|\hat{\rho}^g(n)|^2$. However, the power
spectrum given by the FFT of $\tilde{\rho}^g(x)$ [eq.(45)] is
\begin{equation}
|\hat{\tilde{\rho}}_l^g(n)|^2
 =  \sum_{n'=-\infty}^{\infty}|\hat{W}(n + 2^j n')|^2
|\hat{\rho}^g(n + 2^j n')|^2
\end{equation}
where $\hat{W}(n)$ is the FT of the binning function $W(x)$. The power
spectrum (57) is obviously not equal to the power
spectrum $|\hat{\rho}^g(n)|^2$. The power spectrum (57) is given by a 
superpositions of the power spectrum
$|\hat{\rho}^g(n + 2^j n')|^2$ on all scales $n + 2^j n'$. This is
the ``aliasing'' effect (Hockney \& Eastwood 1989, Hoyle, et al. 1999).

In the DWT representation, the FT of eq.(53) yields
\begin{equation}
\hat{\rho}^g(n)=\sum_{l=0}^{2^J-1}\epsilon^g_{J,l}\hat{\phi}_{J,l}(n)+
 \sum_{j=J}^{\infty}\sum_{l=0}^{2^j-1}
   \tilde{\epsilon}^g_{j,l}\hat{\psi}_{j,l}(n)
\end{equation}
where the function $\hat{\phi}_{j,l}(n)$ is the Fourier transform
of $\phi_{j,l}(x)$, i.e.
\begin{equation}
\hat{\phi}_{j,l}(n) =
\int_{-\infty}^{\infty}\phi_{j,l}(x)e^{-i2\pi nx/L}dx.
\end{equation}
Using the definition of $\phi_{j,l}(x)$ [eq.(A30)], eq.(59)
becomes
\begin{equation}
\hat{\phi}_{j,l}(n)= \left (\frac{2^j}{L} \right )^{-1/2}
 \hat{\phi}(n/2^j)e^{-i2\pi nl/2^j}
\end{equation}
where  $\hat{\phi}(n)$ is the Fourier transform of the basic
scaling function $\phi(\eta)$
\begin{equation}
\hat{\phi}(n)=\int_{-\infty}^{\infty} \phi(\eta) e^{-i2\pi
n\eta}d\eta.
\end{equation}
Eq.(58) gives then
\begin{equation}
\hat{\rho}^g(n)= \left (\frac{2^J}{L} \right )^{-1/2}\hat{\phi}(n/2^J)
\sum_{l=0}^{2^J-1}\epsilon^g_{J,l}e^{-i2\pi n l/2^J}+
\sum_{j=J}^{\infty}\left (\frac{2^j}{L} \right )^{-1/2}
\hat{\phi}(n/2^J)\hat\psi(n/2^j)
\sum_{l=0}^{2^j-1}\tilde{\epsilon}^g_{j,l}e^{-i2\pi nl/2^j},
\end{equation}

Since $\hat\psi(n/2^j)$ is localized in $n/2^j \sim n_p$, the
second terms in the r.h.s. of eq.(62) are important only for $n
\geq 2^Jn_p$. Thus, the Fourier transform $\hat{\rho}^g(n)$ can be
evaluated by
\begin{equation}
\hat{\rho}^g(n) = \hat{\phi}(n/2^J) \hat{F}(n/2^J),  \ \ \
     n \leq 2^J n_p
\end{equation}
where
\begin{equation}
\hat{F}(n/2^J) =\left (\frac{2^J}{L} \right )^{-1/2}
  \sum_{l=0}^{2^J-1}\epsilon_{J,l}e^{-i2\pi n l/2^J}.
\end{equation}
$\hat{F}$ can be calculated by the standard FFT technique.
Therefore, the FT of the galaxy distribution $\rho^g(x)$ can be
evaluated directly by FFT of its SFC mass assignment
$\epsilon^g_{J,l}$. Eqs.(63) and (64) is  actually a
scale-adaptive FFT for estimating the power spectrum of an
irregular data set. This algorithm computes $\hat{\rho}^g(n)$ up
to the scales $n \leq 2^J n_p$, where the adapted  scale $J$ can
be chosen as high as the scales to be studied.

\section{The DWT algorithm on the Poisson sampling}

The observed or the mock galaxy distributions $\rho^g(x)$ are
considered to be a Poisson sampling with an intensity
$\rho^M(x)=\bar{\rho}(x)[1+ \delta(x)]$, where $\bar{\rho}(x)$ is
the galaxy distribution if galaxy clustering is absent, and given
by the selection function (Peebles 1980). A proper
power spectrum estimator should be effective to obtain the power
spectrum debiased from the Poisson sampling. It has been realized
that, to handle the Poisson sampling with a non-uniform
selection function, the decomposition basis $\psi_i({\bf x})$
[eq.(1)] is required to have zero average (e.g. Tegmark et al.
1998), i.e.
\begin{equation}
\int \psi_i(x)dx =0.
\end{equation}
This is what we can take the advantage of the DWT
analysis, as for the wavelets $\psi_{j,l}(x)$, eq.(65) always holds 
due to the admissibility [eq.(7)].

\subsection{Algorithm for the DWT covariance affected by Poisson sampling}

Considering the Poisson sampling, the characteristic function of
the galaxy distribution $\rho^g(x)$ is
\begin{equation}
Z[e^{i\int\rho^g(x)u(x)dx}]= \exp\left \{ \int dx
\rho^M(x)[e^{iu(x)}-1] \right \},
\end{equation}
and the correlation functions of $\rho^g(x)$ are given by
\begin{equation}
\langle\rho^g(x_1)...\rho^g(x_n)\rangle_P =\frac{1}{i^n} \left [
\frac {\delta^n Z}{\delta u(x_1)... \delta u(x_n)} \right ]_{u=0},
\end{equation}
where $\langle ...\rangle_P$ is the average for the Poisson
sampling. We have then
\begin{equation}
\langle\rho^g(x)\rangle_P = \rho^M(x),
\end{equation}
and
\begin{equation}
\langle\rho^g(x)\rho^g(x')\rangle_P = \rho^M(x)\rho^M(x') +
\delta^D(x-x')\rho^M(x).
\end{equation}
This equation yields
\begin{equation}
\langle \delta({\bf x})\delta({\bf x'})\rangle= 1+ \left \langle
\frac{\langle\rho^g(x)\rho^g(x')\rangle_P}
  {\bar{\rho}(x)\bar{\rho}(x') } \right \rangle
- \delta^D(x-x')\frac{1}{\bar{\rho}(x)}.
\end{equation}
Since $\bar{\rho}(x)$ is not subject to a Poisson process, the
second term of the r.h.s. of eq.(70) can be rewritten as $\langle
\langle[\rho^g(x)/\bar{\rho}(x)][\rho^g(x')/\bar{\rho}(x')]
\rangle_P\rangle$. Using eq. (44), we have
\begin{equation}
\frac{\rho^g(x)}{\bar{\rho}(x)} =\sum_{i=1}^{N_g}
\frac{1}{\bar{\rho}(x_i)}w_i\delta^D(x-x_i).
\end{equation}
in which the factor $\bar{\rho}(x_i)$ can be absorbed into the
weight factors $w_i$. The WFC covariance is given by
\begin{equation}
\langle\tilde{\epsilon}_{j,l}\tilde{\epsilon}_{j',l'}\rangle =
\langle \langle\tilde{\epsilon}^g_{j,l}\tilde{\epsilon}^g_{j',l'}
 \rangle_P \rangle
- \int\frac{\psi_{j,l}(x)\psi_{j',l'}(x)}{\bar{\rho}(x)}dx.
\end{equation}
The first term in r.h.s of eq.(70) disappears as all the basis
functions $\psi_{j,l}(x)$ are admissible [eq.(7)].

\subsection{The estimators for the DWT band power spectrums}

If the selection function varies slowly on a scale $j$, i.e.
\begin{equation}
\frac{d\ln \bar{\rho}(x)}{dx} \ll 2^j/L,
\end{equation}
we have approximately,
\begin{equation}
\int\frac{\psi_{j,l}(x)\psi_{j',l'}(x)}{\bar{\rho}(x)}dx=
\frac{1}{\bar{\rho}(x_{l})}\delta_{j,j'}\delta_{l,l'},
\end{equation}
where $\bar{\rho}(x_{l})$ is the number density of galaxies
averaged over a volume of $L/2^j$ at $l$. In this case, the
band-power spectrum is simplified as
\begin{equation}
P_j= \frac{1}{2^j}\sum_{l=0}^{2^j-1} \langle
\langle\tilde{\epsilon}^g_{j,l}\tilde{\epsilon}^g_{j,l} \rangle_P
\rangle -\frac{1}{2^j}
\sum_{l=0}^{2^j-1}\frac{1}{\bar{\rho}(x_{l})}.
\end{equation}
The second term in the r.h.s. is the variance from the Poisson
process. Since the Poisson process does not change the ergodicity,
the average over $l$ in eq.(75) is already a fair estimation for
the ensemble average. Therefore, one can drop $\langle
\langle...\rangle_P \rangle$ in eq.(75), and the estimation of the
DWT band power spectrum is given by
\begin{equation}
P_j= \frac{1}{2^j}\sum_{l=0}^{2^j-1}
\tilde{\epsilon}^g_{j,l}\tilde{\epsilon}^g_{j,l}
 - \frac{1}{2^j}\sum_{l=0}^{2^j-1}\frac{1}{\bar{\rho}(x_{l})}.
\end{equation}
The second term is for subtracting the contribution of the
discreteness effect (or shot noise) in the Poisson sampling from
the power spectrum. $P_j$ is debiased from the Poisson process.

\subsection{The estimators for the scale-scale corrections}

Similarly, one can calculate the debiased scale-scale correlations
from a galaxy sample $\rho^g(x)$. From eq.(70), the term of the Poisson
process is free from scale-scale correlation, the second order
scale-scale correlation can be calculated from the WFCs of the galaxy
distribution without the correction for the shot noise
\begin{equation}
C_{j,j'}(\Delta l) = \frac{1}{2^j} \sum_{l=0}^{2^{j}-1}
 \tilde{\epsilon}^g_{j,l}
\tilde{\epsilon}^g_{j',{\rm mod}[l/2^{j-j'}]+\Delta l}. \ \ \
j>j'.
\end{equation}

However, the Poisson process is not free from higher order scale-scale
correlations. For instance, to estimate the band-band correlations
eq.(42), we use eq.(67) with $n=4$. It gives
\begin{eqnarray}
C^2_{j,j'} &  = & \frac{1}{2^j} \left [ \sum_{l=0}^{2^{j}-1}
(\tilde{\epsilon}^g_{j,l})^2(\tilde{\epsilon}^g_{j',l'})^2  \right
.
   \\ \nonumber
           & & -2\sum_{l=0}^{2^{j}-1}
  \int\frac{\psi_{j,l}(x)\psi_{j',l'}(x)}{\bar{\rho}(x)}dx
  \int\frac{\psi_{j,l}(x')\psi_{j',l'}(x')}{\bar{\rho}(x')}dx' \\ \nonumber
    & & \left . -\sum_{l=0}^{2^{j}-1}
  \int\frac{\psi_{j,l}^2(x)}{\bar{\rho}(x)}dx
  \int\frac{\psi_{j',l'}^2(x')}{\bar{\rho}(x')}dx
-\sum_{l=0}^{2^{j}-1}
\int\frac{\psi^2_{j,l}(x)\psi^2_{j',l'}(x)}{\bar{\rho}^3(x)}dx
\right ].
\end{eqnarray}
where $j>j'$ and $l' = {\rm mod}[l/2^{j-j'}]+\Delta l$. The last
three terms are the scale-scale correlations  $C^2_{j,j'}$ from
the Poisson sampling. Exactly, the factor $\bar{\rho}(x)$ in the
Poisson terms should be
$\rho^M(x)=\bar{\rho}(x)[1+ \delta(x)]$, but we ignored the
contributions of $\delta(x)$ at the moment.

If the selection function is slowly varying on scales $j$ and $j'$
[eq.(73)], we have
\begin{eqnarray}
C^2_{j,j'} &  = & \frac{1}{2^j}\left [ \sum_{l=0}^{2^{j}-1}
(\tilde{\epsilon}^g_{j;l})^2(\tilde{\epsilon}^g_{j';l'})^2 \right
.
  \\ \nonumber
    &  & \left .
    -\sum_{l=0}^{2^{j}-1}\frac{1}{\bar{\rho}(x_{l})\bar{\rho}(x_{l'})}
-\sum_{l=0}^{2^{j}-1}
\int\frac{\psi^2_{j,l}(x)\psi^2_{j',l'}(x)}{\bar{\rho}^3(x)}dx\right
].
\end{eqnarray}
The second and third terms correct for the shot noise on the 4-th
order. Numerical results showed that for typical samples of galaxy
survey the local ($l'=l$) scale-scale correlation of the Poisson
sampling is significant on small scales (Feng, Deng \& Fang
2000.)

\section{Discussions and conclusions}

We presented the method of extracting the band-power spectrum from
observed data and simulation sample via a DWT multiresolution
decomposition. The DWT scale-by-scale approach provides a physical
insight into the covariance matrix of the cosmic mass field.

A key indicator of the DWT power spectrum estimator is the scale-scale
and/or the band-band correlations, which can be calculated directly
from the DWT covariance and the WFCs. In the scale range that the
scale-scale correlations are negligible, the DWT covariance is
$j$(scale)-diagonal, and it is already a lossless estimation of a
banded power spectrum $P_j$. This DWT band power spectrum is
optimized in the sense that the spatial resolution is adaptive
automatically to the scales of the density perturbations.

In the scale range that the scale-scale (or band-band)
correlations are significant, the diagonalization of the
covariance may not yield an accurate power spectrum, but seriously
contaminated by errors. In this case, an effective confrontation
between the observed sample and model-prediction may not be given
by a full diagonalized covariance, but both of the DWT power
spectrum and scale-scale correlations. With the DWT
representation, one can calculate the scale-scale correlation as
well as the DWT power spectrum. Therefore, the DWT covariance is
also useful when scale-scale correlation is strong.

In summary, the basic DWT algorithm is proceeded in the following
steps,
\begin{enumerate}
\item Calculation of the WFCs $\tilde{\epsilon}^g_{J,l}$ and/or the SFCs
      $\epsilon^g_{J,l}$ from the data $\rho^g(x)$, where $J$
      corresponds to the highest resolution of the samples.
\item Calculation of the WFCs $\tilde{\epsilon}^g_{j,l}$ for various scale
      $j$.
\item Calculate the band-power spectrum $P_j$, and scale-scale correlations
      $C_{j,j'}$.
\item In the $j$ range of $C_{j,j'}\simeq 0$, testing models or
      constraining parameters by comparing the model-predicted DWT
      band-power spectrum $P_j$ with observed results.
\item In the $j$ range of $C_{j,j'}\neq 0$, testing model or constraining
     parameters by comparing the model-predicted DWT band-power spectrum
     and scale-scale correlations with observed results.
\end{enumerate}

Since the DWT is computationally powerful, the above-mentioned algorithm
is found to be numerically efficient and flexiable (Yang et al. 2000.) 
Moreover, the developed method is open in the sense
that based on the WFCs and SFCs one can add subsequent items to
realize the further goals related to the power spectrum measurement
and model discrimination. Some of these problems are discussed below.

\subsection{Higher dimensions and complex geometry}

The DWT analysis in a 2 and/or 3-D space ${\bf x}$ can be
performed by the bases of the 1-D bases direct product, i.e.
\begin{equation}
\psi_{(j_1, j_2,j_3),(l_1,l_2,l_3)}(x_1,x_2,x_3) =
 \psi_{j_1,l_1}(x_1)\psi_{j_2,l_2}(x_2)\psi_{j_3,l_3}(x_3).
\end{equation}
In this case, the three scales $(j_1,j_2,j_3)$ of the WFCs can be
different for different directions. One can define radial scales
by
\begin{equation}
k = 2\pi \left [ \left (\frac{2^{j_1}}{L_1}\right )^2 +
        \left (\frac{2^{j_2}}{L_2}\right )^2 +
        \left (\frac{2^{j_3}}{L_3}\right )^2
                       \right ]^{1/2}  \ ,
\end {equation}
where $L_1 \times L_2 \times \times L_3$ is the 3-D box.

For 2 and 3-D samples, one can also decompose by the mixed direct
product of 1-D wavelets and scaling functions. For instance, a 3-D
sample can be decomposed by bases
\begin{equation}
\psi^{(1,2)}_{(j_1,j_2,j_3),(l_1,l_2,l_3)}(x_1, x_2, x_3)=
    \phi_{j_1,l_1}(x_1)\psi_{j_2,l_2}(x_2)\psi_{j_3,l_3}(x_3).
\end{equation}
where the scaling functions $\phi_{j,l}$ actually play the role of
chopping a 3-D sample into $2^{j_1}$ 2-D slices in the $x_1$
direction, $l_1 = 0, ... 2^{j_1}-1.$ Like the binning by the
scaling function (\S 4.2), the chopping eq.(82) will not cause
spurious features.

The problem of complex geometry of samples can be treated by using
the locality of the $\psi_{j,l}$ (Pando \& Fang 1998a). The
locality property allows the WFCs to be independent of the data
outside an ``influence" cone. The WFCs $\tilde{\epsilon}_{j,l}$ is
only determined by data in the interval $[(lL/2^{j+1} - (\Delta
x)/2^{j+1}, (lL/2^{j+1} + (\Delta x)/2^{j+1}]$, where $\Delta x$
is the width of the basic wavelet $\psi$. With this property, any
complex geometry of samples can be regularized into a 2 or 3-D box
by zero padding in the field between the sample geometry and the
box. Since all WFCs at the zero padding zone are zero, one can use
the DWT to analyze the regular box, but not treat the WFCs related
to the zero padding as the variables of valid degrees of
freedom.\footnote{About
DWT on manifold, see also W. Sweldens http://cm.bell-labs.com/who/wim
or http://www.wavelet.org}

\subsection{Non-Gaussianity and power spectrum detection}

We have emphasized that the information of the non-Gaussian
features are important for a precise detection of the power
spectrum, or band power spectrum. That is because, from the covariance,
one can only find statistically uncorrelated (or statistical orthogonal)
bases or modes on second order.  For non-Gaussian fields,
the modes statistically uncorrelated on second order might be
statistically correlated at the 3rd and 4th orders. On the other hand,
the power spectrum is of second order, and therefore, the power spectrum
estimates at different scales might not be statistically uncorrelated
if there are 3rd and 4th order correlations. The accuracy of a power
spectrum estimation is affected by the higher order statistical
correlations.

For instance, a popular bias model for galaxy formation employ the
selection probability functions as (Cole et al. 1998)
\begin{equation}
P(\delta({\bf r})) \propto
  \exp \left [\alpha \frac{\delta_s({\bf r})}{\sigma_s}\right ],
\end{equation}
where $\alpha$ is const, and $\delta_s({\bf r})$ and $\sigma_s$
are smoothed density field and variance. Therefore, if the density
field is Gaussian, the galaxy distribution given by the Poisson
sampling with the intensity eq.(83) will be lognormal. The
baryonic distribution is sometimes also modeled by a lognormal
relation with the underlying Gaussian mass field (Bi, Ge \& Fang
1995, Bi \& Davidsen 1997). As having been well known, for
lognormal distribution, the most likely value can be significantly
different from their mean value. In this case, to estimate the
accuracy of a power spectrum detection, the higher order cumulant
statistics is needed.

In the DWT analysis, the $2^j$ WFCs give the one point
distribution of the fluctuations on scale $j$. Therefore, the
third and forth cumulants can be calculated by
\begin{equation}
S_j = \frac{1}{P_j^{3/2}} \frac{1}{2^j} \sum_{l=0}^{2^j-1}
(\tilde{\epsilon}_{j,l} - \overline{\tilde{\epsilon}_{j,l}})^3. ,
\end{equation}
\begin{equation}
K_j  \equiv   \frac{1}{P_j^2} \frac{1}{2^j} \sum_{l=0}^{2^j-1}
(\tilde{\epsilon}_{j,l} - \overline{\tilde{\epsilon}_{j,l}})^4 -3
\end{equation}
These are, respectively, the  skewness and kurtosis spectra. It is
not difficult to generalize eqs.(84) and (85) to more higher
orders.

\subsection{Selection of the basis of the multiresolution analysis}

In computing the samples of redshift surveys, there are two
coordinate systems having been widely used: 1. parallel plane
system; 2. spherical shell system. For system 1, the volume of the
survey can be approximated as a box, and therefore, the wavelets
of eqs.(80) and (82) are suitable for the decomposition. For the
system 2, we should use the wavelets on 2-D spherical surface.
With the development of the DWT analysis, the bank of the DWT
analysis has stored more and more sets of the orthogonal and
complete basis for the multiresolution decomposition of different
geometries. The multiscale analysis on geometry beyond
above-mention two simple cases is being feasible.

\subsection{Systematic effects}

The influence of various systematic effects on the power spectrum
detection has only been studied very preliminarily. The linear
effect of redshift distortion on the power spectrum detection has
been well studied (e.g. Hamilton 1995). It is not difficult to
incorporate the linear theory of the redshift distortion with the
DWT analysis. A key operator of the mapping a real space
distribution into redshift space is $(1-a (\partial^2/\partial
z^2)\nabla^{-2})$, where coefficient $a$ is const. To diagonalize
this differential-integral operator, the Fourier representation is
certainly the best. However, it has been shown that this operator
is quasidiagonal in the DWT representation (Farge 1996).

Moreover, it would be straightforward to include a scale-dependent
bias in the DWT representation. The redshift distortion is usually
calculated under the assumption that the galaxy distribution
$\rho^g({\bf x})$ is linearly related to the underlying mass field
$\rho({\bf x})$, i.e. $\rho^g({\bf r})=b \rho({\bf r})$, where $b$
is the bias parameter. However, observations have indicated that
the bias parameters probably are scale-dependent (Fang, Deng \&
Xia 1998.) It is easy to introduce scale-dependent bias in the DWT
representation. For instance one can define a bias parameter on
scale by $\tilde{\epsilon}^g_{j,l}=b_j\tilde{\epsilon}_{j,l}$.

\acknowledgments

LLF acknowledges support from the National Science Foundation of
China (NSFC) and World Laboratory Scholarship. This project was
done during LLF's visiting to the Department of Physics,
University of Arizona. This work was supported in part by the LWL
foundation.  We thank anonymous referee for helpful comments.

\appendix

\section{The discrete wavelet transform (DWT) of density fields}

Let us briefly introduce the DWT analysis of the cosmic mass
density fields, for the details of mathematical stuffs refers to
the classical papers by Mallat (1989a,b,c); Meyer (1992);
Daubechies, (1992) and references therein, and for physical
applications, refers to Fang \& Thews (1998) and references
therein. Some other cosmological applications of wavelets can also
be found at, e.g., Pando, Vills-Gabaud \& Fang (1998), Hobson,
Jones \& Lasenby (1999), Sanz et al. (1999), Tenorio et al.
(1999), Xu, Fang, \& Wu (2000), Cayon, et al (2000).

\subsection{Expansion by scaling functions}

We consider here a 1-D mass density distribution $\rho(x)$ or
contrast $\delta(x)=[\rho(x)-\bar{\rho}]/\bar{\rho}$, which are
mathematically random fields over a spatial range $0 \leq x \leq
L$. It is not difficult to extend all results developed in this
section into 2-D and 3-D because the DWT bases for higher
dimension can be constructed by a direct product of 1-D bases.

First, we introduce the scaling functions for the Haar wavelets.
There are top-hat window functions defined by 
\begin{equation}
\phi_{j,l}^{H}(x) = \left\{ \begin{array}{ll} 1 & \mbox{for
$Ll2^{-j} \leq x \leq L(l + 1)2^{-j}$}\\ 0 & \mbox{otherwise.}
\end{array} \right. ,
\end{equation}
where the superscript $H$ is stand for Haar. The scaling function,
$\phi_{j,l}^{H}(x)$ actually gives a window at
resolution scale $L/2^j$ and position $Ll2^{-j} \leq x \leq L(l +
1) 2^{-j}$. With the scaling function, the mean of density
contrast distribution in the spatial range $Ll2^{-j} \leq x \leq
L(l + 1) 2^{-j}$ can be expressed as
\begin{equation}
\epsilon_{j,l}=
\frac{2^j}{L}\int_{0}^{L}\delta(x)\phi_{j,l}^{H}(x)dx.
\end{equation}
The number $\epsilon_{j,l}$ is called the scaling function
coefficient(SFC). Using SFCs, one can construct a density contrast
field as
\begin{equation}
\delta^{j}(x) = \sum_{l=0}^{2^{j}-1}\epsilon_{j,l}
\phi_{j,l}^{H}(x).
\end{equation}
This is the density contrast $\delta(x)$ smoothed on scale
$L/2^j$, or for simple, $j$-scale.

The scaling function $\phi_{j,l}^{H}(x)$ can be rewritten
\begin{equation}
\phi_{j,l}^{H}(x)= \phi^{H}(2^{j} x/L - l),
\end{equation}
where
\begin{equation}
\phi^{H}(\eta) = \left\{ \begin{array}{ll} 1 & \mbox{for 0 $\leq
\eta \leq$ 1}\\ 0 & \mbox{otherwise.}
\end{array} \right.
\end{equation}
$j$, $l$ are integers, with $j \geq 0$, and $0 \leq l \leq 2^j -
1$. $\phi^{H}(\eta)$ is called the basic scaling function. The
scaling function $\phi_{j,l}^{H}(x)$ is thus a translation and 
dilation of the basic scaling function.

The functions $\phi_{j,l}^{H}(x)$ are orthogonal with respect to
$l$, i.e.
\begin{equation}
\int_{0}^{L}\phi_{j,l}^{H}(x)\phi_{j,l'}^{H}(x)dx =
\frac{L}{2^j}\delta_{l,l'}
\end{equation}
where $\delta_{l,l'}$ is Kronecker delta function. Thus, eq.(A3)
gives functions in the function space $V_j$ spanned by bases
$\phi_{j,l}^{H}(x)$. $V_j$ is a closed subspaces of $L_2(R)$, i.e.
$V_j \subset L_2(R)$. It is easy to show that
\begin{equation}
\phi_{j,l}^{H}(x)=\phi^{H}_{j+1,2l}(x) + \phi^{H}_{j+1,2l+1}(x)
\end{equation}
\begin{equation}
\epsilon_{j,l} = \frac {1}{2} (\epsilon_{j+1,2l} +
\epsilon_{j+1,2l+1}).
\end{equation}
Therefore, $V_j \subset V_{j+1}$ for all $j$. Thus, the orthogonal
projectors $P_j$ onto $V_j$, i.e. $P_j f \in V_j$, satisfy
\begin{equation}
\lim_{j\rightarrow \infty} P_jf=f,
\end{equation}
for all $f\in L_2(R)$. A multiresolution analysis is then defined
by the sequence of subspaces $V_j$.

\subsection{Expansion by wavelets}

Eqs. (A7) and (A8) show that $\delta^j(x)$ contains less information 
than $\delta^{j+1}(x)$, because information on scale $j+1$ have been
smoothed out by eq. (A8). It would be nice not to lose any
information during the smoothing from $j+1$ to $j$ [eq.(A8)]. This can be
accomplished if the differences,  $\delta^{j+1}(x)-\delta^{j}(x)$,
between the smoothed distributions on succeeding scales are
somehow retained. This is, if we are able to retain these {\it
differences}, this scheme will then make it possible to smooth the
distribution and yet not lose any information as a result of the
smoothing.

To calculate the {\it differences}, we define the difference
function, or wavelet, as
\begin{equation}
\psi^{H}(\eta) = \left\{ \begin{array}{ll} 1 & \mbox{for $0 \leq
\eta \leq 1/2$} \\ -1 & \mbox{for $1/2 \leq \eta \leq 1$} \\ 0 &
\mbox{otherwise.}
\end{array} \right.
\end{equation}
This is the basic Haar wavelet. As with the scaling functions, one
can construct a set of wavelets $\psi_{j,l}^{H}(x)$ by dilating
and translating eq.(A10) as
\begin{equation}
\psi_{j,l}^{H}(x)  =
   \psi^{H}(2^{j} x/L - l).
\end{equation}
The Haar wavelets are orthogonal with respect to {\it both}
indexes $j$ and $l$, i.e.
\begin{equation}
\int_0^L \psi_{j',l'}^{H}(x)\psi_{j,l}^{H}(x)dx =
\left(\frac{L}{2^j}\right)\delta_{j',j} \delta_{l',l} .
\end{equation}
For a given $j$, $\psi_{j,l}^{H}(x)$ is also orthogonal to the
scaling functions $\phi_{j',l}^{H}(x)$ with $j'\leq j$, i.e.
\begin{equation}
\int_0^L \phi_{j',l'}^{H}(x)\psi_{j,l}^{H}(x)dx = 0, \hspace{2cm}
{\rm if \ \ \ } j' \leq j.
\end{equation}

 From eqs.(A4) and (A11), we have
\begin{equation}
\begin{array}{ll}
\phi_{j,2l}^{H}(x) &
    = \frac{1}{2} (\phi_{j-1,l}^{H}(x) + \psi_{j-1,l}^{H}(x)),\\
      &   \\
\phi_{j, 2l+1}^{H}(x) &
   = \frac{1}{2}(\phi_{j-1,l}^{H}(x) - \psi_{j-1,l}^{H}(x)).\\
\end{array}
\end{equation}
Thus, the difference $\delta^{j+1}(x)-\delta^{j}(x)$ is given by
\begin{equation}
\delta^{j+1}(x)-\delta^{j}(x)=
 \sum_{l=0}^{2^{j}-1} \tilde{\epsilon}_{j,l} \psi_{j-1,l}^{H}(x),
\end{equation}
where $\tilde{\epsilon}_{J-1,l}$ are called the wavelet function
coefficients(WFC), which is given by
\begin{equation}
\tilde{\epsilon}_{j,l}=\frac{2^{j}}{L}
 \int \delta(x)\psi_{j,l}^{H}(x)dx .
\end{equation}

Using the relation (A15) repeatedly, we have
\begin{equation}
\delta^{j}(x) = \delta^{0}(x)+
     \sum_{j'=0}^{j-1} \sum_{l=0}^{2^{j'}-1} \tilde{\epsilon}_{j',l}
 \psi_{j',l}^{H}(x).
\end{equation}
This is an expansion of the function $\delta^{j}(x)$ with respect
to the basis $\psi_{j,l}^{H}(x)$, and $\delta^{0}(x)$ is the mean
of $\delta(x)$ in the range $L$. We have $\delta^{0}(x)=0$ if
$\delta(x)$ is density contrast. Considering (A9), for any $f(x)
\in L^2(R)$ in $L$ with mean $\bar{f}=0$ we have
\begin{equation}
f(x)= \sum_{j=0}^{\infty} \sum_{l=0}^{2^{j}-1}
\tilde{\epsilon}_{j,l}
 \psi_{j,l}^{H}(x),
\end{equation}
and
\begin{equation}
\tilde{\epsilon}_{j,l} = \frac{2^{j}}{L} \int_0^L
f(x)\psi_{j,l}^{H} dx.
\end{equation}

For a given $j$, the wavelets $\psi^H_{j,l}(x)$ form a space
$W_{j}$ which is the orthogonal complements of $V_{j}$ in
$V_{j+1}$, i.e. $V_{j+1}=V_{j} \oplus W_{j}$. Thus, every $f^j \in
V_j$ has a unique decomposition $f^j = f^{j-1} + d^{j-1}$ with
$f^{j-1} \in V_{j-1}$ and $d^{j-1} \in W_{j-1}$. Since $W_j
\subset V_{j+1}$ and $W_j$ is orthogonal to $V_{j}$, $W_j$ is also
orthogonal to $W_{j-1}$ and $W_{j+1}$. Thus, all the spaces $W_j$
are mutually orthogonal. Since $V_j$ contains only $W_{j'}$ with
$j' < j$, $V_j$ is orthogonal to all $W_{j'}$ with $j'\geq j$.

\subsection{Compactly supported orthogonal basis}

In terms of the subspace $V_j$, the basic scaling function
$\phi(\eta)$ and basic $\psi(\eta)$ belong to $V_0$ and $W_0$
respectively, and they can be expressed by the basis of $V_1$,
$\phi(2\eta - l)$, i.e.
\begin{equation}
\phi(\eta)  = \sum_{l=-\infty}^{\infty} a_l \phi(2\eta-l), \\
\nonumber
\end{equation}
\begin{equation}
\psi(\eta)  = \sum_{l=-\infty}^{\infty} b_{l} \phi(2\eta - l),
\end{equation}
where $a_l$ and $b_l$ are called the filter coefficients.

If we require that the scaling function $\phi(\eta)$ is
normalized, eq.(A21) yields
\begin{equation}
\sum_l a_l= 2.
\end{equation}
Requiring orthogonality for $\phi(x)$ with respect to discrete
integer translations, i.e.
\begin{equation}
\int_{-\infty}^{\infty} \phi(\eta-m) \phi(\eta) d\eta =
\delta_{m,0},
\end{equation}
we have
\begin{equation}
\sum_l a_l a_{l+2m} = 2 \delta_{0,m}.
\end{equation}
The orthogonality between $\phi$ and $\psi$ means
\begin{equation}
\int_{-\infty}^{\infty} \psi(\eta)\phi(\eta -l) d\eta = 0.
\end{equation}
Therefore, one has
\begin{equation}
b_l=(-1)^la_{1-l}.
\end{equation}
Furthermore, the wavelet $\psi(\eta)$ has to be admissible
\begin{equation}
\int_{-\infty}^{+\infty}\psi(\eta) d\eta = 0,
\end{equation}
so we need
\begin{equation}
\sum_l b_l = 0.
\end{equation}

The conditions (A22), (A24), (A26) and (A28) for the filter
coefficients were employed to construct families of scaling
functions and wavelets. The simplest solution of the filter
coefficients is $a_0=a_1=b_0=-b_1=1$ and all others 0. This
solution gives the Haar wavelet. After the Haar wavelet, the
simplest solution for the filter coefficients is
\begin{eqnarray}
a_0=(1+\sqrt 3)/4, & \  a_1=(3+\sqrt 3)/4, \\ \nonumber
a_2=(3-\sqrt 3)/4, & \  a_3=(1-\sqrt 3)/4.
\end{eqnarray}
This is the Daubechies 4 wavelet (D4). It is compactly supported
and continuous.

With these wavelets, the multiresolution analysis can be performed
in the similar way as developed in last two sections for the Haar
wavelets. The scaling functions and wavelets for spanning the subspace
$V_j$ and $W_j$ are given, respectively, by a translation and dilation
of the basic scaling function and basic wavelet
\begin{equation}
\phi_{j,l}(x) = \left (\frac{2^{j}}{L}\right)^{1/2} \phi(2^jx/L -
l)
\end{equation}
and
\begin{equation}
\psi_{j,l}(x) = \left (\frac{2^{j}}{L}\right)^{1/2}\psi(2^jx/L-l).
\end{equation}

The wavelets are orthonormal, i.e.
\begin{equation}
\int \psi_{j,l}(x)\psi_{j',l'}(x)dx=\delta_{j,j'} \delta_{l,l'}.
\end{equation}
Eqs.(A23) and (A25) yield also
\begin{equation}
\int \phi_{j,l}(x)\phi_{j,l'}(x)dx= \delta_{l,l'},
\end{equation}
and
\begin{equation}
\int \phi_{j,l}(x)\psi_{j',l'}(x)dx= 0 \hspace{1cm} j' \geq j.
\end{equation}
The set of $\psi_{j,l}$ and
 $\phi_{0,m}(x)$ with $ 0 \leq j < \infty$
and $- \infty < l, m < \infty$ form a complete, orthonormal basis
in the space of functions with period length $L$.

Thus, a density field $\rho(x)$ with period length $L$ can be
expanded as (Fang \& Thews 1998)
\begin{equation}
\rho(x) = \bar\rho + \bar\rho
 \sum_{j=0}^{\infty} \sum_{l= - \infty}^{\infty}
  \tilde{\epsilon}_{j,l} \psi_{j,l}(x),
\end{equation}
or the density contrast $\delta(x)=(\rho(x)-\bar\rho)/\bar\rho$ is
\begin{equation}
\delta(x) = \sum_{j=0}^{\infty} \sum_{l= - \infty}^{\infty}
  \tilde{\epsilon}_{j,l} \psi_{j,l}(x),
\end{equation}
where
\begin{equation}
\bar\rho =L^{-1}\int_0^L \rho(x)dx
\end{equation}
and
\begin{equation}
\tilde{\epsilon}_{j,l}=\int_{-\infty}^{\infty} \delta(x)
\psi_{j,l}(x)dx.
\end{equation}

More generally, we have
\begin{equation}
\rho(x) =\rho^J(x)+
  \bar{\rho}\sum_{j=J}^{\infty}\sum_{l=-\infty}^{+\infty}
      \tilde{\epsilon}_{j,l}\psi_{j,l}(x),
\end{equation}
where $\rho^{J}(x)$ is the density field smoothed on scale $J$
\begin{equation}
\rho^J(x)=\sum_{l=-\infty}^{+\infty}\epsilon_{J,l}\phi_{J,l}(x).
\end{equation}
and the scaling function coefficient(SFC) $\epsilon_{J,l}$ is
given by
\begin{equation}
\epsilon_{J,l} =\int_{-\infty}^{+\infty}\rho(x)\phi_{J,l}(x)dx.
\end{equation}

\end{document}